\documentclass[11pt]{article}
\usepackage{spconf,amsmath,epsfig,color}
\usepackage[dvipsnames]{xcolor}
\colorlet{LightRubineRed}{RubineRed!70!}
\colorlet{Mycolor1}{OliveGreen!90!}
\definecolor{Mycolor2}{HTML}{00F9DE}
\usepackage{flushend,cite}
\usepackage[hyphens]{url}
\usepackage{upgreek}

\usepackage[utf8]{inputenc}
\usepackage[T1]{fontenc}
\usepackage{amsmath}
\usepackage{graphicx} 
\usepackage{hyperref}
\usepackage{subfig}
\usepackage[export]{adjustbox}

\usepackage{xcolor}
\usepackage[thickqspace,cdot,squaren]{SIunits}

\title{Application of DinSar technique to high coherence satellite images for strategic infrastructure monitoring}

\name{\small{${\it T.\ De\ Corso}^1$, ${L.\ Mignone}^1$, ${A.\ Sebastianelli}^1$, ${M.P.\ Del\ Rosso}^1$, ${C.\ Yost}^2$,  ${E.\ Ciampa}^1$, ${M.\ Pecce}^1$}, ${S.\ Sica}^1$, ${S.\ Ullo}^1$} 
\address{$^{(1)}$ University of Sannio, Benevento (Italy), {\it tonydecorso3@gmail.com,luca1297.lm@gmail.com, }\\
{\it alessandro.sebastianelli1995@gmail.com, mariapia.delrosso@gmail.com, eciampa@unisannio.it}\\
{\it pecce@unisannio.it,  stefsica@unisannio.it, ullo@unisannio.it}\\
$^{(2)}$ Massachusetts Institute of Technology (MIT), USA, {\it cyost@mit.edu }}

\begin{document}
\maketitle 
\begin{abstract}
In this paper  the authors present and validate a procedure, which intends to  combine the latest state of the art models in bridge monitoring with freely available satellite data.
Through the Differential SAR interfer\nobreak ometry (DinSAR) technique, a dataset of displacements for the Morandi bridge  in Genoa (Italy), before its collapse, has been created, by using images downloaded by the Copernicus Open-Access Hub and the ASFVer\nobreak tex Hub.
The data have been processed through the ESA SNAP software to identify the rate of displacements in the parts of the bridge where collapse occurred. Results demonstrate 
that  the adopted procedure has great potentiality in the application field, as it represents a simple and inex\nobreak pensive method to monitor large structures in a continuous way, by helping  to better quantify risks and guide effective mitigation countermeasures.
 Moreover, the same procedure, once properly validated, could be effectively extended to the current and future performance estimation  of   civil infrastructures. 
\end{abstract}
\begin{keywords}
Bridge monitoring, Differential Synthetic Aperture Radar Interferometry (DInSAR), Sentintel-1, Copernicus Open-Access, ASFVertex
\end{keywords}

\vspace{-0.4cm}

\section{Introduction}
\label{sec:intro}
\vspace{-0.3cm}
Aging and deterioration of structures such as bridge, dams or other strategic infrastructures are becoming an urgent social issue due to the huge economic and human loss related to their collapse. Structures in fact could mo\nobreak dify their behaviour, during their service life, due to deterioration of materials, changes in loads, environmental conditions, design and constructions errors \cite{Pecce2019}. Nowadays there is an increasing need of quantifying the current and future performance of existing structures by means of advanced structural health monitoring and damage detection tools \cite{chang2003health} \cite{ciampa2019practical}. Several monitoring methods with different features may be adopted in different situations \cite{sousa2014potential}.     The contact methods, which requires the use of accelerometers and fiber-optic, could monitor the most critical points of the construction in a very short time interval, but they cannot be adopted to monitor the whole bridge because they are time-consuming and expensive  \cite{sousa2014potential}. Traditional methods, unfortunally, based on a limited set of sensors mounted on the bridge, collect point-like information and have the disadvantage of providing incomplete (in space and time) displacement information \cite{huang2017displacement}. Even though satellite image processing techniques do not allow distinguishing between stress accumulation or material degradation processes, they could be fundamental in detecting important structural deficiencies and anomalous behaviours. The combined use of remote sensing imagery and on-site data could in fact strengthen the quality of the overall monitoring process.
In this paper the authors applied a multidisciplinary approach comparing and validating the outputs obtained through the DInSAR technique to a more complex and reliable technique called Persistent Scatterer Interferometry (PSI).  It is worth emphasizing that with this approach it is possible to detect structural displacements of the order of millimeters. 

\begin{figure*}[!ht]
	\centering
	\includegraphics[scale=0.6]{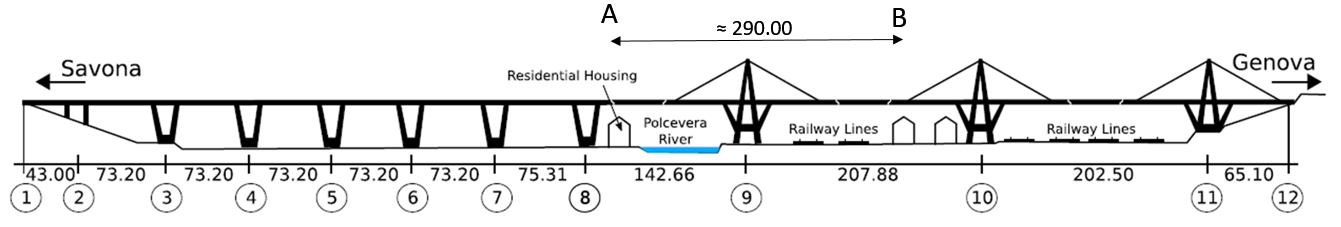}
 \caption{Two-dimensional structural scheme of the bridge, from West (1) to East (12). Dimensions in meters \cite{calvi2019once}.} 
	\label{The Morandi Bridge}
\end{figure*}

\section{The Morandi Bridge}
The Morandi Bridge, also called Polcevera viaduct for the name of the river underneath it, was designed by Riccardo Morandi and  built in the 1960s \cite{calvi2019once},  \cite{morandi1979long},  \cite{morandi1967viadotto}. 
It is one of the Italian  most important bridges with prestressed concrete stays \cite{martinez1994rehabilitation}. This bridge connected the motorways Milan-Genoa and Genoa-Savona, crossing a heavily urbanized area and rising above civil and industrial buildings. The bridge was a prestressed concrete cable-stayed bridge composed of 11 spans, for a total length of about 1102 meters, supported by two single piers (1 to 2), six V-shaped piers (3 to 8) consisting of two reinforced concrete inclined piers, and A-shaped frames consisting of three independent balanced systems (see Figure \ref{The Morandi Bridge}). The three-span continuous deck were supported by stay cables, made of steel and concrete shells casted around the cables \cite{morandi1967viadotto}. The function of these shells, besides protecting the steel cables, was to reduce the cable elongation during the passage of traffic loads because of prestressing of the shells themselves. The deck was entirely made of reinforced concrete. 

\begin{figure}[!ht]
	\centering
	\includegraphics[scale=0.3]{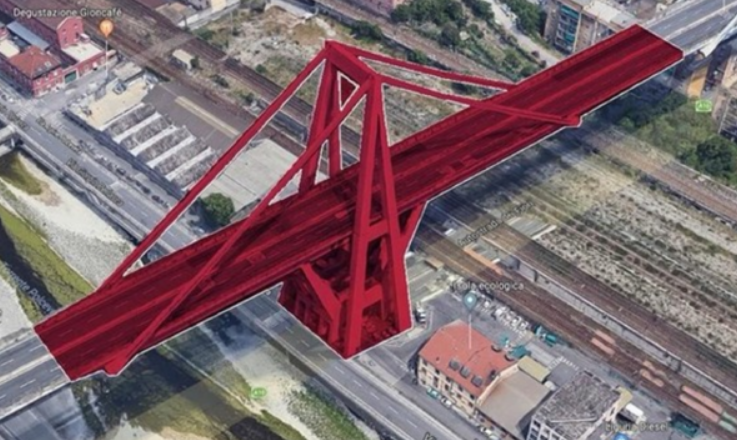}
	\caption{Collapsed stack 9 balanced system highlighted in red  \cite{monodaq}}
	\label{The Morandi Bridge_2}
\end{figure}

On August 14, 2018, one of the three balanced system (the number 9 in the Figure \ref{The Morandi Bridge}) collapsed, causing the death of 43 people. The collapsed balanced system is highlighted in red in the Figure \ref{The Morandi Bridge_2}.

In the 1990s, as reported in \cite{calvi2019once} and  \cite{martinez1994rehabilitation}: “during maintenance and repair activities, it was discovered that the stays of the three balanced systems were suffering from widespread general deterioration, as well as several instances of concentrated degradation”. The structure, moreover, was exposed to marine winds canalized in the valley, although the sea is about 1 km away, and struck directly by acid fumes from the chimneys.
The progressive reduction of the steel section and of the compressed concrete section in the stays could have led to increased local displacements and deck level irregularities \cite{calvi2019once}. 
\section{THE CASE STUDY}
\begin{figure}[!ht]
    \centering
    \includegraphics[scale=0.4]{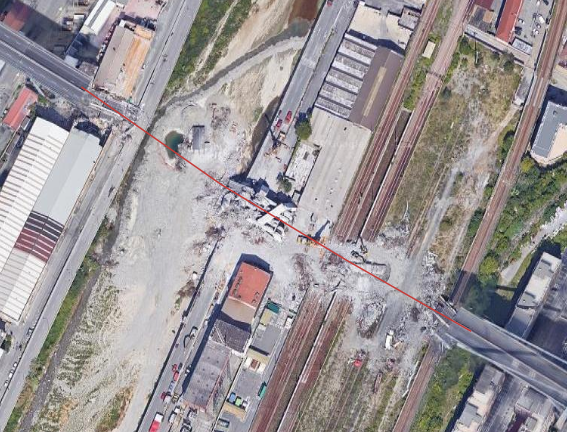}
    \caption{The single line covering the stack 9 balanced system: only measures on this line have been considered.}
    \label{vector_data}
\end{figure}

 Analysis has been carried out from August 2017 to August 2018, until a few days before the bridge collapsed. In particular, a single line covering the stack 9 balanced system (see Figure \ref{vector_data}) has been considered, along which   the time-series displacement map for a certain number of selected pixels was built.

\begin{figure}[!ht]
    \centering
    \includegraphics[scale=0.3]{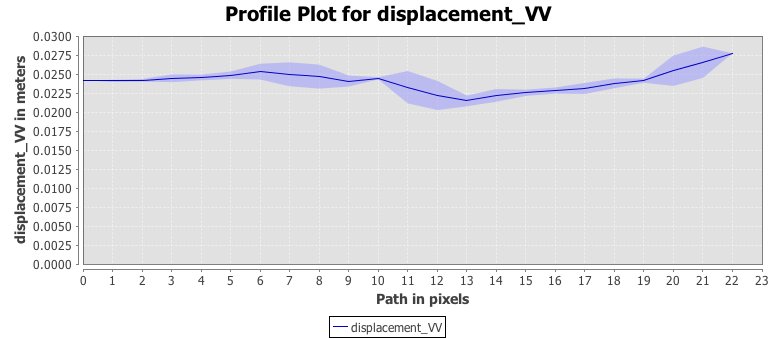}
    \caption{Displacements along the bridge from 08-08-2017 to 08-20-2017}
    \label{profile_image}
\end{figure}
\begin{figure}[!ht]
    \centering
    \includegraphics[scale=0.50]{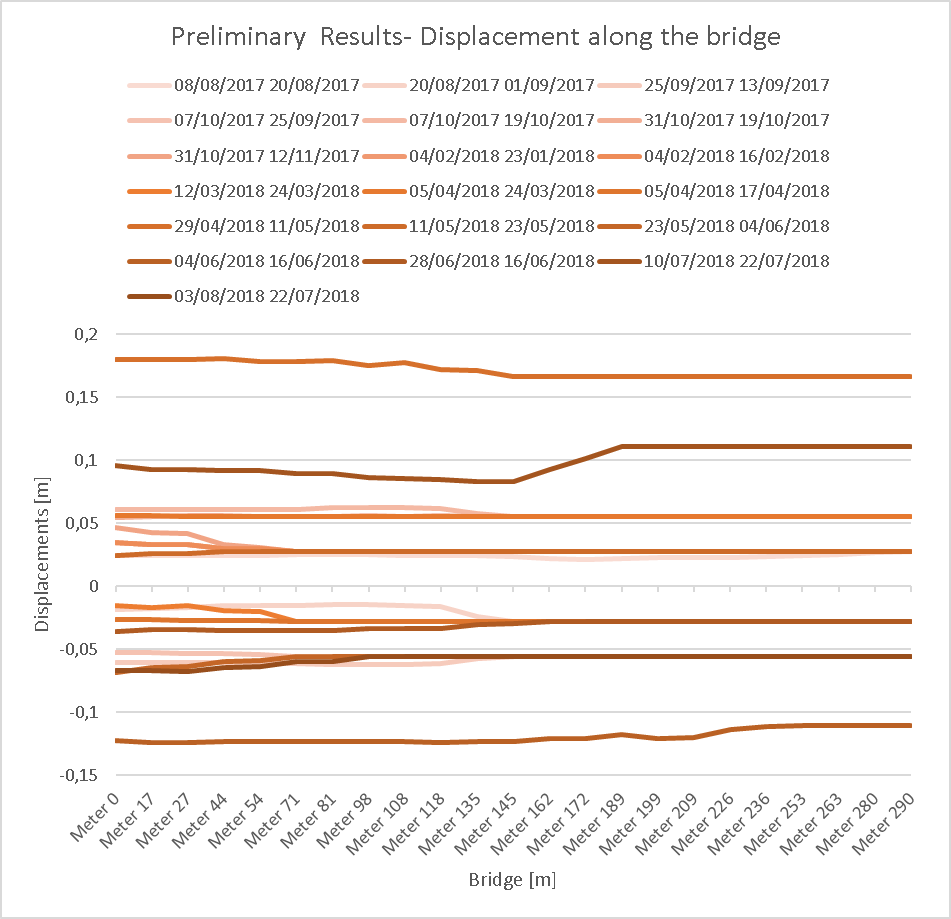}
    \caption{Preliminary results - Detected displacements along the bridge}
    \label{plot_1}
\end{figure}
\begin{figure}[!ht]
    \centering
    \includegraphics[scale=0.76]{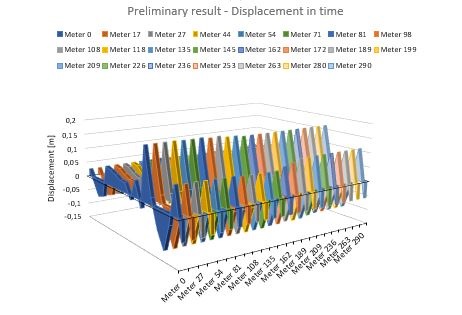}
    \caption{Preliminary results - Detected displacements in time}
    \label{plot_2}
\end{figure}

\vspace{-0.2cm}
For the pair of images $08/08/2017-08/20/2017$, in the Figure \ref{profile_image} the displacements are shown once the Profile Plot tool of SNAP has been applied. 
This information has then been acquired for the whole sequence of master-slave pairs of  images, over the one-year mon\nobreak itor\nobreak ing period, by obtaining a set of vectors describing how the points on the bridge were moving overtime, as shown in the  Figure \ref{plot_1}. For the sake of clarity, it is worth to clarify that  geographical coordinates have been converted into distances between pixels, therefore the two Figures \ref{profile_image} and \ref{plot_1} display number of pix\nobreak els in one case, and meters in the other. 
Moreover, in the Figure \ref{plot_2} a different perspective of the previous results is shown.  This work is in progress and detailed description of the methodology and discussion of results will be presented in the final paper. Yet, it is already possible to highlight how the portion was subjected to stress over the one-year period.
It is worth to say that the difference of displacements of a point of the deck near the antenna, during the observed period,  was in agreement with the results reported in  \cite{milillo2019pre}. This  corroborates the reliability of the proposed methodology.
 
Subsequent uses of the methodology could be automating this process, in order to obtain a deformation trend  for a timely notice against a possible collapse, by showing the potential operational use of the DInSAR technique for the health monitoring of critical infrastructures.
\vspace{-0.2cm}
\section{Procedure applied}
In this paper, the authors tried to extend the work done in \cite{ullo2019application}, where a procedure was proposed and validated through the combination of \emph{in situ} measurements and the use of free Sentinel-1 images and the free ESA SNAP toolbox. 

The idea has been to exploit the use  of freely available data, accessible from different sources, like  the  Copernicus Open-Access Hub by ESA \cite{esaopen} and the ASF Vertex by NASA \cite{asfvertex}. Moreover 
a more advanced technique has been considered as a reference, the PSI (Persistent Scatterer Interferometry), for evaluation of results obtained through the DInSAR  \cite{crosetto2016persistent}, \cite{delgado2019measuring}. The peculiarity of PSI is that this technique does not consider the whole image but only high coherence points, adding reliability and robustness to the results. Since infrastructures are supposed to be fixed in time, they represent a natural source of high coherence points, and therefore the PSI well  fits applications in the field of infrastructure monitoring.  

\newpage
 A time-series of interferograms could be extremely useful for structure monitoring, as they could offer a fast way to read anomalous displacements remotely, covering a large area of interest with high resolution. 
 Nevertheless, bridge monitoring through remote sensing techniques such as DInSAR or PSI can be challenging, since some pixels on the ground can interfere with pixels on the bridge, resulting in meaningless information. To avoid these effects, a large quantity of data needs to be acquired, and at this end it turns very useful that the SNAP software supports Python plugins, so it can be programmed for downloading and processing huge numbers of SAR images, as already available plugins do \cite{delgadoautomated}.\\
 This work is in progress and comparison between DInSAR and PSI will be discussed in the final paper. 
  \section{Conclusions and future works}
 In this paper freely available satellite data have been processed through the ESA SNAP software  to  identify  the  rate  of  displacements  in  the parts  of  the Morandi bridge  where  collapse  occurred.   Results demonstrate that the adopted procedure has great potentiality in the application field, as it represents a simple and inexpensive method to monitor large structures in a continuous way. 
 Future works will employ the SARPROZ software \cite{perissin2014sarproz} to validate the procedure in \cite{ullo2019application} not only for ground movements, but also for infrastructure ones.
 


\footnotesize
\bibliographystyle{IEEEbib}
\bibliography{refs}
\end{document}